\documentclass[lettersize,journal]{IEEEtran}
\usepackage{amsmath}
\usepackage{amssymb}
\usepackage{times}
\usepackage{graphicx}
\usepackage[colorlinks=true, allcolors=blue]{hyperref}
\usepackage{booktabs}
\usepackage{array}
\usepackage{makecell}
\usepackage{pifont}
\usepackage{caption}
\usepackage{authblk}
\usepackage[T1]{fontenc}
\usepackage{multirow}
\usepackage{hyperref}
\usepackage{longtable}    
\usepackage{tabularx}
\usepackage{booktabs}
\usepackage{pifont}   
\usepackage{array}    
\usepackage{comment}
\usepackage{subcaption} 
\usepackage{microtype}  
\usepackage[table]{xcolor}
\usepackage{tabularx}
\usepackage{booktabs}

\usepackage{booktabs}      
\usepackage{tabularx}      
\usepackage{ragged2e}      
\usepackage{makecell}      
\usepackage{caption}       

\usepackage{tikz}

\UseRawInputEncoding

\usepackage{makecell}
\setcellgapes{4pt}
\makegapedcells

\usepackage[english]{babel}

\AtBeginDocument{%
  \fontdimen2\font=0.21em 
  \fontdimen3\font=0.1em  
  \fontdimen4\font=0.06em  
}

\begin{document}
\bstctlcite{IEEEexample:BSTcontrol}
\title{Scene-Aware Latency Estimation for Microservices via Multi-Scale Graph Fusion} 

\author{
        Zhichao~Sun,
        Hailiang~Zhao,~\IEEEmembership{Member,~IEEE},
        ~and~
        Kingsum~Chow
\thanks{
    Zhichao Sun, Hailiang Zhao, and Kingsum Chow are with the School of Software Technology, Zhejiang University. Emails: \{sunzc, hliangzhao, kingsum.chow\}@zju.edu.cn.}
}

\maketitle

\begin{abstract}
Cloud-Native microservice architectures have become prevalent owing to their inherent flexibility and scalability properties. To satisfy service quality guarantees, cloud providers must implement efficient proactive autoscaling algorithms. 
However, effective proactive scaling critically depends on accurately estimating end-to-end latency under given resource quotas, which remains highly challenging. 
Existing methods struggle with the multi-hierarchical nature and dynamic operational contexts of microservice systems. 
They primarily employ single-scale modeling that fails to capture inherent organizational structures and lacks adaptability to varying workload types.
To address these limitations, we propose MSGAF, a Multi-Scale Graph Adaptive Fusion framework with Scene-Aware Learning for microservice latency estimation. 
Our approach constructs hierarchical graph representations through learnable aggregation-based coarsening, capturing system behaviors across microscopic, mesoscopic, and macroscopic levels. 
The framework comprises three components: a system state encoding module transforming heterogeneous monitoring data into unified representations, a multi-scale graph adaptive fusion module leveraging graph attention networks for hierarchical feature extraction, and a scene-aware learning module employing specialized expert networks with dynamic weight allocation for context-specific estimation. 
Additionally, we design and implement a comprehensive non-intrusive monitoring system for real-time data collection.
Extensive experiments on benchmark microservice applications demonstrate that MSGAF significantly outperforms state-of-the-art methods across diverse operational scenarios, providing substantial improvements for cloud-native performance optimization.
\end{abstract}

\section{Introduction} \label{sec:intro}
\begin{figure}[htbp]
    \centering
    \includegraphics[width=0.98\columnwidth]{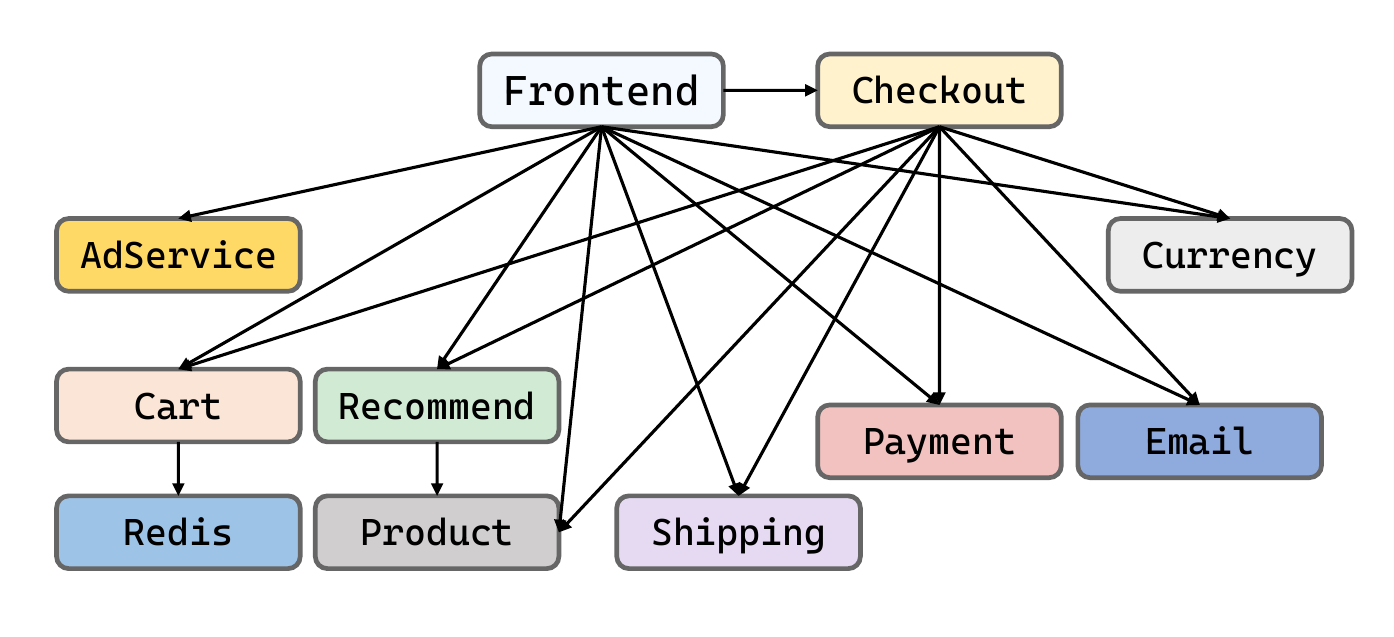}
    \caption{Microservices and their internal call relations in Online Boutique.}
    \label{fig:online-boutique-arch}
\end{figure}

Microservice architecture has already become a dominant paradigm in modern cloud-native applications, breaking down monolithic systems into loosely coupled, independently deployable microservices \cite{dragoni2017microservices}. This modular design offers significant advantages in terms of flexibility, scalability, and fault isolation, enabling organizations to rapidly iterate and scale their services dynamically. As illustrated in Figure~\ref{fig:online-boutique-arch}, a representative example such as the Online Boutique application \cite{google_microservices_demo} demonstrates the complex interdependencies among microservices, where each service performs a specific function and communicates with others through well-defined APIs.

However, the inherent complexity and dynamic nature of microservice systems pose significant challenges for performance management. Fluctuating workloads and intricate service interactions can lead to unpredictable system behavior, increasing the risk of Service Level Agreement (SLA) violations \cite{liu2018elasticity,alharthi2024auto}. To mitigate such risks, cloud providers often resort to \textit{over-provisioning} resources, which, although effective, results in suboptimal resource utilization and increased operational costs. Hence, achieving a balance between SLA compliance and resource efficiency has become a critical objective in microservice resource management.

One promising approach to address this challenge is \textit{autoscaling}, which dynamically adjusts resource allocation in response to workload variations. Autoscaling strategies are typically categorized into two types: reactive and proactive \cite{al2024containerized}. Reactive approaches allocate resources based on real-time system metrics (e.g., CPU, memory, and I/O utilization) \cite{gias2019atom,kannan2019grandslam}, while proactive methods leverage predictive models to anticipate future demands and optimize resource allocation accordingly \cite{rzadca2020autopilot,chow2022deeprest,nguyen2022graph}. The latter has gained increasing attention due to its potential to reduce resource waste and improve system responsiveness. At the core of effective proactive autoscaling lies the ability to \textit{accurately estimate end-to-end service latency} under varying resource configurations. However, achieving precise latency estimation in microservice systems remains a challenging task due to their complex and dynamic nature. Current approaches to latency estimation can be broadly classified into three categories: (i) \textit{model-driven methods} that rely on analytical performance models \cite{gias2019atom,bhasi2021kraken}, (ii) \textit{data-driven black-box models} using machine learning techniques \cite{zhang2021sinan,gan2019seer}, and (iii) \textit{graph-aware hybrid methods} that incorporate system structure into learning frameworks \cite{chow2022deeprest,park2024graph,nguyen2022graph}.

Despite notable progress, these approaches suffer from two key limitations when applied to modern microservice environments: First, most existing methods adopt a single-scale modeling paradigm, failing to capture the hierarchical organization of microservice systems. Actually, microservices naturally form a multi-level structure, in which fine-grained components at the lower layer, functionally grouped modules at the middle layer, and end-to-end user-facing services at the top layer. Our analysis of real-world benchmarks such as Online Boutique \cite{google_microservices_demo} and Sock Shop \cite{holbach2022sockshop} reveals that these hierarchical relationships are often implicit and cannot be easily extracted from configuration files or manual inspection. Second, microservice systems exhibit high runtime variability, operating under diverse scenarios such as CPU-bound, I/O-bound, and network-bound workloads. The dominant performance bottlenecks and contributing factors to latency differ significantly across these scenarios. As a result, a one-size-fits-all estimation model often fails to generalize across different operational contexts.



To address these challenges, we propose a framework called MSGAF (\underline{M}ulti-\underline{S}cale \underline{G}raph \underline{A}daptive \underline{F}usion with Scene-Aware learning) for microservice latency estimation. MSGAF integrates a multi-scale graph representation learning architecture with a scene-aware adaptive inference mechanism, enabling accurate and robust latency prediction across diverse operational conditions. Specifically, our key contributions are as follows:


\begin{itemize}
    \item We introduce a multi-scale graph representation learning framework that captures the hierarchical structure of microservice systems. By constructing a layered graph abstraction and performing adaptive fusion across scales, our method overcomes the limitations of existing single-scale modeling approaches.
    
    \item We propose a scene-aware adaptive mechanism that dynamically identifies operational patterns at runtime and adjusts the estimation strategy accordingly. This allows the model to maintain high accuracy across different workload scenarios, including CPU-, I/O-, and network-intensive conditions.


    \item We implement a non-intrusive performance monitoring system based on service mesh techniques (e.g., Istio\footnote{\texttt{https://istio.io/}} and Prometheus\footnote{\texttt{https://prometheus.io/}}), enabling real-time data collection without modifying application code. Extensive experiments on two widely adopted benchmarks, i.e., Online Boutique~\cite{google_microservices_demo} and Sock Shop~\cite{holbach2022sockshop}, demonstrate that MSGAF outperforms state-of-the-art methods in terms of latency estimation accuracy under real-world trace-driven simulations.
\end{itemize}

\section{Related Work}\label{sec:related}

Latency estimation in microservice architectures has been approached through various methodologies, which can be broadly categorized into model-driven, data-driven, and hybrid approaches.

\subsection{Model-Driven Approaches}

Model-driven methods construct theoretical frameworks to represent microservice system behavior and predict performance characteristics. ATOM~\cite{gias2019atom} builds upon Layered Queuing Networks (LQN) to model service interactions and predict component-level workloads, which are then aggregated for system-wide performance estimation. Kraken~\cite{bhasi2021kraken} adopts Variable Order Markov Models (VOMM) to capture temporal patterns in service invocations and predict future performance based on historical invocation sequences. Erms~\cite{luo2022erms} employs a piecewise linear function to model and analyze individual microservice latency characteristics.

\subsection{Data-Driven Approaches}

Data-driven methodologies treat microservice systems as black-box entities and employ machine learning techniques to learn performance patterns directly from operational data. Sinan~\cite{zhang2021sinan} integrates Convolutional Neural Networks (CNNs) with Boosted Trees to establish direct mappings between workload characteristics and end-to-end latency. Seer~\cite{gan2019seer} combines CNN architectures with Long Short-Term Memory networks to capture both spatial and temporal patterns in performance data. However, these approaches typically ignore structural dependencies between components and require extensive training data, limiting their interpretability and applicability under novel conditions.

\subsection{Graph-Aware Hybrid Approaches}

To address the limitations of both model-driven and data-driven methods, hybrid approaches have emerged that combine structural modeling with machine learning techniques. Sage~\cite{gan2021sage} employs Causal Bayesian Networks to model component dependencies and utilizes Graphical Variational Auto-Encoders to simulate performance under various scenarios. FIRM~\cite{qiu2020firm} introduces critical path analysis combined with Reinforcement Learning to optimize performance along identified bottleneck paths.

\begin{figure*}[!h]
  \centering
  \includegraphics[width=0.95\linewidth]{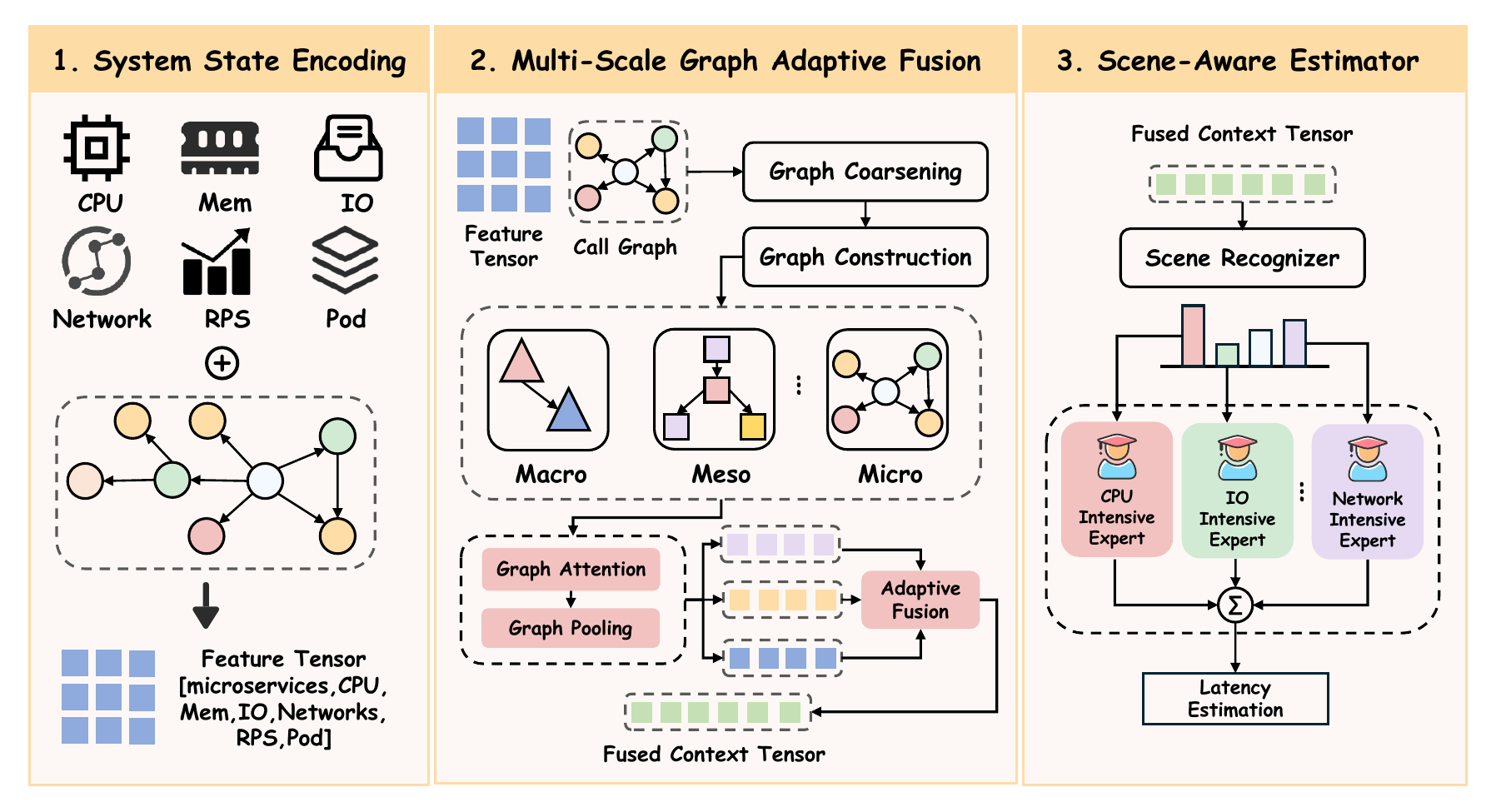}
  \caption{The architecture of the MSGAF framework. It consists of three modules: 1) System State Encoding Module; 2) Multi-Scale Graph Adaptive Fusion Module; and 3) Scene-Aware Estimation Module.}
  \label{fig:overall-model}
\end{figure*}

Recently, Graph Neural Networks (GNNs) have gained attention for microservice performance modeling due to their ability to capture both structural relationships and learn complex patterns from data. DeepRest~\cite{chow2022deeprest} processes span graphs to capture request execution flows and provide API-level performance predictions. GRAF~\cite{park2024graph} leverages graph node embeddings with Message Passing Neural Networks (MPNN) to predict application-level latency metrics. PERT-GNN~\cite{nguyen2022graph} builds PERT graphs from execution traces, preserving temporal orderings while applying GNNs for accurate latency prediction.

\section{Problem Formulation}

In microservice systems, end-to-end latency is affected by multiple factors including resource quota, workload distribution, inter-service call dependencies, and dynamic runtime scenes. 

We model the microservice call dependency as a directed graph $\mathcal{G} = (\mathcal{V}, \mathcal{E}, \mathbf{A})$, where $\mathcal{V}$ represents the set of microservice nodes with $|\mathcal{V}| = n$, $\mathcal{E}$ denotes the edge set of inter-service calls, and $\mathbf{A} \in \mathbb{R}^{n \times n}$ is the adjacency matrix where $A_{ij} = 1$ if and only if there exists a call relationship from service $v_i$ to service $v_j$.
Given the resource quota vector $\mathbf{C} \in \mathbb{R}^{n \times 1}$, observed system state metrics $\mathbf{S} \in \mathbb{R}^{n \times d}$, and workload request vector $\mathbf{W} \in \mathbb{R}^{n \times 1}$, where the system state metrics include key performance indicators such as CPU utilization, memory utilization, and network traffic, we aim to construct function $f_\theta$ for end-to-end latency estimation at time $t$:
\begin{equation}
\hat{L} = f \big( \theta; \mathcal{G}, \mathbf{S}, \mathbf{C}, \mathbf{W} \big),
\end{equation}
where $\hat{L} \in \mathbb{R}^+$ represents the estimated end-to-end latency, $\theta$ denotes the model parameter set.

\section{Methodology}


\subsection{Model Architecture}

The architecture of the MSGAF framework is shown in Figure~\ref{fig:overall-model}. It follows an end-to-end deep learning paradigm that effectively captures both the structural dependencies and multi-scale behavioral patterns of microservice systems. The model operates through a structured pipeline consisting of five stages: feature encoding, multi-scale graph representation, adaptive fusion, scene-aware adaptation, and latency estimation.

\subsection{System State Encoding}

The system state encoding module serves as the foundational component of the entire estimation model, with the core task of transforming multi-dimensional heterogeneous system monitoring data into unified feature representations.

This module's inputs include system state metrics $\mathbf{S} \in \mathbb{R}^{n \times d}$, resource quota vector $\mathbf{C} \in \mathbb{R}^{n \times 1}$, and workload request rate $\mathbf{W} \in \mathbb{R}^{n \times 1}$. In selecting system state metrics, we are guided by the understanding that end-to-end latency in microservice systems arises from the propagation of requests through service call chains. Therefore, our feature selection follows the principle of bottleneck identification, prioritizing metrics that have the most direct impact on service responsiveness and system performance. \textit{CPU utilization} and \textit{memory utilization} are chosen as key indicators of service processing capacity. These metrics become dominant contributors to latency when resources approach saturation, significantly affecting service execution time. Beyond compute-bound factors, \textit{file I/O metrics} are included to capture disk-bound bottlenecks, which are particularly critical for stateful services requiring persistent storage access. Given the distributed nature of microservice architectures, \textit{network traffic} plays a central role in performance modeling. Specifically, \textit{incoming network traffic} reflects upstream call pressure, while \textit{outgoing network traffic} reveals downstream dependencies. These metrics are essential for understanding communication overhead, fault propagation, and cascading delays in distributed environments. Based on this analysis, we select the following five key performance indicators: CPU utilization, memory utilization, file I/O metrics, incoming network traffic, and outgoing network traffic. In addition to performance metrics, we incorporate two contextual features to capture system configuration and workload dynamics. \textit{Pod count} is selected as a representation of deployment scale, as it directly influences the system’s concurrent processing capacity. For workload intensity, we use \textit{calls per minute} to quantify the real-time request pressure experienced by each service, enabling dynamic modeling of load variations. These metrics are concatenated with resource quota $\mathbf{C}$ and workload intensity $\mathbf{W}$ into a unified feature matrix:
\begin{equation}
    \mathbf{X} = \big[\mathbf{S}, \mathbf{C}, \mathbf{W}\big] \in \mathbb{R}^{n \times (d+2)}.
\end{equation}

\subsection{Multi-Scale Graph Adaptive Fusion}

The multi-scale graph adaptive fusion module is the core component of the MSGAF framework, designed to capture hierarchical system characteristics through multi-level graph representations. Given that microservice systems exhibit distinct behavioral patterns across different abstraction levels, single-scale modeling approaches often fail to fully characterize the system's complexity. To address this limitation, our module integrates learnable graph coarsening with attention-based feature learning and an adaptive fusion mechanism, enabling effective extraction and integration of multi-granularity features.

\paragraph{Learnable Aggregation-based Graph Coarsening.}
Traditional graph coarsening methods typically rely on fixed heuristics or domain-specific rules, limiting their adaptability to diverse system states. In contrast, we propose a learnable aggregation strategy that dynamically adapts to node features and topological structure.

The aggregation assignment matrix $\mathbf{P}$ is generated by mapping node embeddings into soft aggregation assignments:
\begin{equation}
    \mathbf{P} = \text{Softmax}\Big(\mathbf{W}_p \mathbf{X} + \mathbf{b}_p\Big) \in \mathbb{R}^{n \times k},
\end{equation}
where $n$ is the number of nodes in the original graph, $k$ is the target number of aggregated nodes, $\mathbf{X} \in \mathbb{R}^{n \times (d+2)}$ denotes the input node feature matrix from the system state encoding module, and $\mathbf{W}_p \in \mathbb{R}^{(d+2) \times k}$ and $\mathbf{b}_p \in \mathbb{R}^{k}$ are learnable parameters. This formulation ensures that aggregation decisions are informed by both semantic similarity and structural connectivity.

\paragraph{Hierarchical Graph Representation Construction.}
Using the learned assignment matrix $\mathbf{P}$, we constructed coarsened graph representation by performing weighted aggregation over node features and adjacency matrices. The coarsened node features are computed by
\begin{equation}
    \mathbf{X}_c = \mathbf{P}^\textrm{T} \mathbf{X}.
\end{equation}
Correspondingly, the coarsened adjacency matrix is
\begin{equation}
    \mathbf{A}_c = \mathbf{P}^\textrm{T} \mathbf{A} \mathbf{P},
\end{equation}


We then construct coarsened graph $\mathcal{G}_c$ from the aggregated representations $\mathbf{X}_c$ and $\mathbf{A}_c$. By setting the number of aggregated nodes $k \in \{n, \lfloor n/4 \rfloor, \lfloor n/8 \rfloor\}$, we obtain three levels of graph representations: (1) \textit{Microscopic level} ($k=n$): it maintains original node granularity, capturing fine-grained inter-service interaction patterns; (2) \textit{Mesoscopic level} ($k=\lfloor n/4 \rfloor$): it aggregates related services into service groups, reflecting medium-scale system behavioral characteristics; (3) \textit{Macroscopic level} ($k=\lfloor n/8 \rfloor$): it describes system load distribution and performance characteristics from a global perspective.

\paragraph{Graph Attention Network for Feature Learning.}
To extract high-level representations at each scale, we employ graph attention networks (GATs), which dynamically assign importance weights to neighboring nodes based on both their features and topological relationships.

To enhance the expressive power of node features and enable more effective attention computation, we apply a learnable linear embedding transformation. This mapping projects the input features into a high-dimensional semantic space, where latent relationships between nodes can be better captured for attention-based aggregation. Specifically, we define the transformation as follows:
\begin{equation}
    \mathbf{H} = \mathbf{X_c} \mathbf{W}^\textrm{T} \in \mathbb{R}^{n \times h},
\end{equation}
where $\mathbf{H}$ is the resulting node embedding matrix.

Given a node $i$ in the coarsened graph, its attention coefficient with respect to neighbor $j$ is computed as
\begin{equation}
    e_{ij} = \text{LeakyReLU}\Big(\mathbf{a}^\textrm{T} \big[\mathbf{W}\mathbf{h}_i ; \mathbf{W}\mathbf{h}_j\big]\Big),
\end{equation}
where $\mathbf{h}_i \in \mathbb{R}^{h}$ denotes the feature vector of node $i$ from the resulting node embedding matrix $\mathbf{H}$, $\mathbf{a} \in \mathbb{R}^{2d'}$ is a trainable attention vector, $\mathbf{W} \in \mathbb{R}^{d' \times h}$ is the linear transformation matrix for attention computation. Then, the attention coefficients are normalized using the softmax function:
\begin{equation}
\alpha_{ij} = \frac{\exp(e_{ij})}{\sum_{k \in \mathcal{N}_i} \exp(e_{ik})},
\end{equation}
ensuring that $\sum_{j \in \mathcal{N}_i} \alpha_{ij} = 1$, where $\mathcal{N}_i$ denotes the neighborhood of node $i$. Node features are then updated via an attention-weighted aggregation:
\begin{equation}
    \mathbf{h}_i' = \sigma\bigg(\sum_{j \in \mathcal{N}_i} \alpha_{ij} \mathbf{W} \mathbf{h}_j\bigg),
\end{equation}
where $\sigma(\cdot)$ is a nonlinear activation function. Finally, we derive a global graph-level representation using mean pooling:
\begin{equation}
    \mathbf{h} = \frac{1}{|\mathcal{V}|}\sum_{i \in \mathcal{V}} \mathbf{h}_i'.
\end{equation}
Applying this process independently to each scale level $l \in \mathcal{L} := \{ \text{micro}, \text{meso}, \text{macro} \}$ yields a set of graph embeddings, each capturing system characteristics at a distinct level of abstraction.


\paragraph{Adaptive Multi-Scale Feature Fusion.}

To effectively integrate complementary information from multiple granularities, we design an learnable adaptive fusion mechanism that dynamically adjusts the contribution of each scale according to the input context. Specifically, the fusion weight $\beta_l$ for each scale level $l \in \mathcal{L}$ is computed as:
\begin{equation}
    \beta_l = \frac{\exp\big(\mathbf{W}_{\beta} \mathbf{h}_l + b_\beta\big)}{\sum_{l' \in \mathcal{L}} \exp\big(\mathbf{W}_{\beta} \mathbf{h}_{l'} + b_{\beta}\big)},
\end{equation}
where $\mathbf{W}_{\beta}$ and $b_{\beta}$ are learnable parameters.
The final fused representation is obtained via a weighted combination:
\begin{equation}
    \mathbf{f} = \sum_{l \in \mathcal{L}} \beta_l \cdot \mathbf{h}_l \in \mathbb{R}^d.
\end{equation}


\subsection{Scene-Aware Estimation}

The scene-aware estimation module adapts its estimation strategy based on the current system state. This module comprises three core components:  a scene recognizer, a dynamic weight generator, and a multi-expert estimator.

The scene recognizer is responsible for identifying the current operational mode of the system based on the fused feature representation $\mathbf{f}$ obtained from the multi-scale graph adaptive fusion module. It employs a two-layer fully connected neural network to extract high-level scene features:
\begin{align}
    \mathbf{s}^{(1)} &= \text{ReLU} \Big( \mathbf{W}_s^{(1)} \mathbf{f} + \mathbf{b}_s^{(1)} \Big), \\
    \mathbf{s} &= \mathbf{W}_s^{(2)} \mathbf{s}^{(1)} + \mathbf{b}_s^{(2)},
\end{align}
where $\mathbf{s} \in \mathbb{R}^{d_s}$ denotes the final scene feature vector that encodes the semantic representation of the current system state. Here, $ \mathbf{W}_s^{(1)}$, $\mathbf{W}_s^{(2)}$ are learnable weight matrices, and $\mathbf{b}_s^{(1)}$, $\mathbf{b}_s^{(2)}$ are bias terms.

Based on the extracted scene features $\mathbf{s}$, the dynamic weight generator computes a set of adaptive weights $\boldsymbol{\omega} = [\omega_1, ..., \omega_K]$, which determine the contribution of each expert in the multi-expert estimator:
\begin{equation}
\boldsymbol{\omega} = \text{Softmax}(\mathbf{W}_\omega \mathbf{s} + \mathbf{b}_\omega).
\end{equation}
Here, $\mathbf{W}_\omega$ and $\mathbf{b}_\omega$ are learnable parameters. This enables the model to automatically prioritize the most relevant expert for the current scene.

The multi-expert estimator consists of $K$ specialized neural networks, each trained to handle a distinct type of operational scene (e.g., CPU-intensive, I/O-intensive, Network-intensive, and Mixed Load). Each expert network $E_i$ is implemented as a two-layer feed-forward network. 
Specifically, for any input $\mathbf{x}$,
\begin{equation}
    E_i(\mathbf{x}) = \text{ReLU}\bigg(\mathbf{W}_i^{(2)} \text{ReLU}\Big(\mathbf{W}_i^{(1)} \mathbf{x} + \mathbf{b}_i^{(1)}\Big) + \mathbf{b}_i^{(2)}\bigg),
\end{equation}
where $\mathbf{W}_i(1)$, $\mathbf{W}_i(2)$, $\mathbf{b}_i^{(1)}$, and $\mathbf{b}_i^{(2)}$ are learnable parameters. The final estimation result $\hat{L}$ is obtained by performing a weighted combination of all expert outputs: $\hat{L} = \sum_{i=1}^{K} \omega_i \cdot E_i(\mathbf{f})$.

\subsection{Loss Function}

Our goal is to minimize prediction error while encouraging functional specialization among expert networks. Thus, the overall loss function consists of two components: the main task loss minimizes the mean squared error between predicted and actual latencies: $\frac{1}{N} \sum_{i=1}^{N} (\hat{L}_i - L_i)^2$, and a KL-divergence-based regularization term encourages expert specialization: $-\frac{1}{K(K-1)} \sum_{i=1}^{K} \sum_{j=1, j \neq i}^K \text{KL}(P_i \| P_j)$. 

\section{System Design and Implementation}

Traditional observability solutions typically rely on trace-based or log-based approaches, such as distributed tracing using Jaeger~\cite{jaeger2025} or Zipkin~\cite{zipkin2025}, and log aggregation with tools like the ELK Stack~\cite{elkstack2025} or Fluentd~\cite{fluentd2025}. However, these methods are often intrusive, requiring the integration of specific SDKs or agents into application code. This leads to tight coupling with business logic and hinders the development of a unified and standardized monitoring system across heterogeneous microservice environments.

\begin{figure*}[htbp]
    \centering
    \includegraphics[width=1.2\columnwidth]{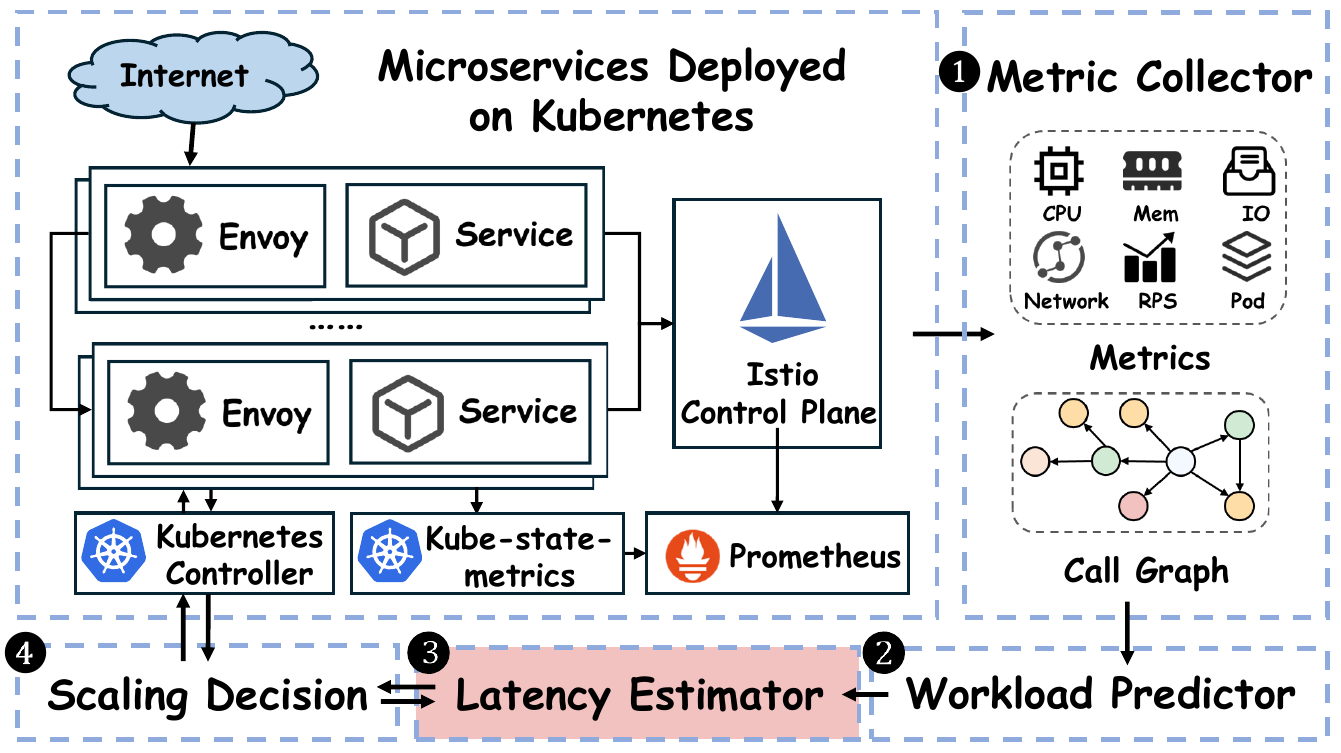}
    \caption{Overview of the non-intrusive system performance monitoring and auto-scaling framework.}
    \label{fig:system}
\end{figure*}

To overcome these limitations, we design and implement a non-intrusive monitoring system, as illustrated in Figure~\ref{fig:system}. In this system, Service Mesh acts as an infrastructure-layer control plane for service-to-service communication. We adopt Istio for use. Istio transparently captures rich data, i.e., request latencies, error rates, and traffic patterns, without requiring any changes to the application code by deploying lightweight sidecar proxies alongside each microservice instance. The collected data is then fed into Prometheus, a time-series monitoring engine that aggregates, stores, and evaluates metrics in real time. Prometheus continuously scrapes metric endpoints exposed by the service mesh and other system components, computes high-level performance indicators such as latency percentiles and success rates, and enables alerting based on predefined thresholds. After that, Kube-state-metrics\footnote{https://github.com/kubernetes/kube-state-metrics} monitors the state of the Kubernetes cluster itself by observing the Kubernetes API server. It exposes metrics about the desired and current states of deployments, pods, and nodes, allowing the system to detect scheduling issues, resource bottlenecks, or failed replicas.


The latency estimator acts as an important module within the auto-scaling framework, as depicted in Figure~\ref{fig:system}. The system architecture comprises four essential modules: \ding{182} Metrics Collector, \ding{183} Workload Predictor, \ding{184} Latency Estimator, and \ding{185} Scaling Decision module. The latency estimator receives predicted workload patterns from the Workload Predictor and integrates them with real-time metrics from the monitoring system. It then estimates system latency under specific resource quotas to provide the Scaling Decision module with performance assessments for various resource allocation scenarios.

\section{Experiments}

\subsection{Experimental Setup}

\subsubsection{Benchmarks.} To evaluate our approach, we use two widely adopted open-source microservice benchmarks that represent diverse application domains and architectural complexities:

\begin{itemize}
    \item {Online Boutique}~\cite{google_microservices_demo} is a cloud-native microservice application developed by Google to showcase Kubernetes, Istio, and gRPC technologies. It simulates an e-commerce platform with features such as product browsing, cart management, and purchase processing, offering a realistic representation of modern microservice-based applications.

    \item {Sock Shop}~\cite{holbach2022sockshop} is a microservice-based e-commerce platform that simulates a sock retail store. Comprising 13 services implemented using Spring Boot, Go Kit, and Node.js, it demonstrates common microservice architecture patterns and cloud-native deployment practices.
\end{itemize}

\subsubsection{Cluster Setup.} All experiments are conducted in a private cloud cluster, consisting of seven machines (one master node and six worker nodes) with a total of 80 vCPUs and 128 GB RAM. All compute nodes are equipped with Intel Xeon E5-2673v4 CPUs.

\subsubsection{Workload.} 
\begin{figure}[tbp]
  \centering
  \begin{subfigure}[b]{0.26\textwidth}
    \centering
    \includegraphics[width=\textwidth]{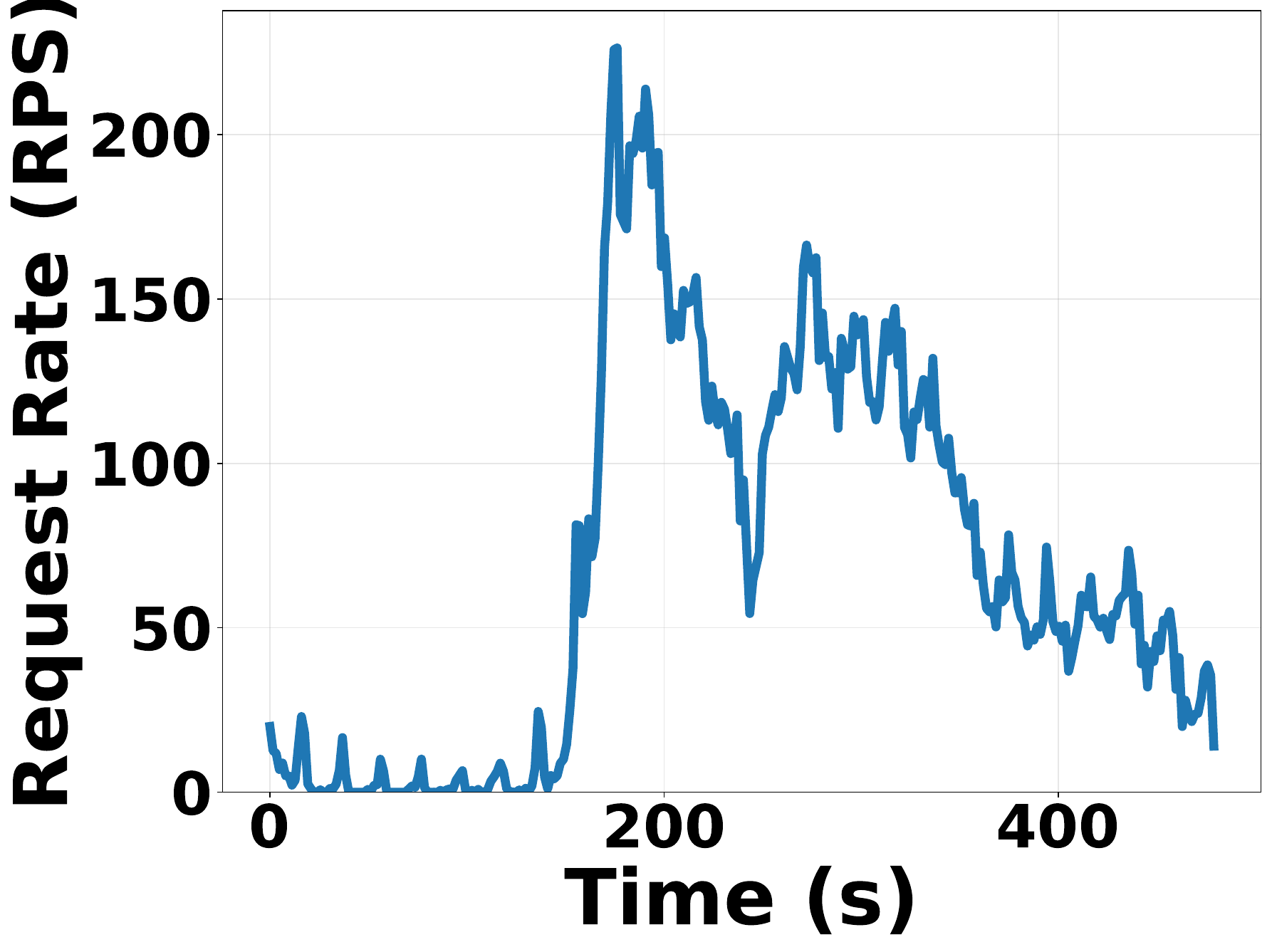}
    \caption{Large fluctuations}
    \label{fig:rps-2}
  \end{subfigure}
  \quad
  \begin{subfigure}[b]{0.26\textwidth}
    \centering
    \includegraphics[width=\textwidth]{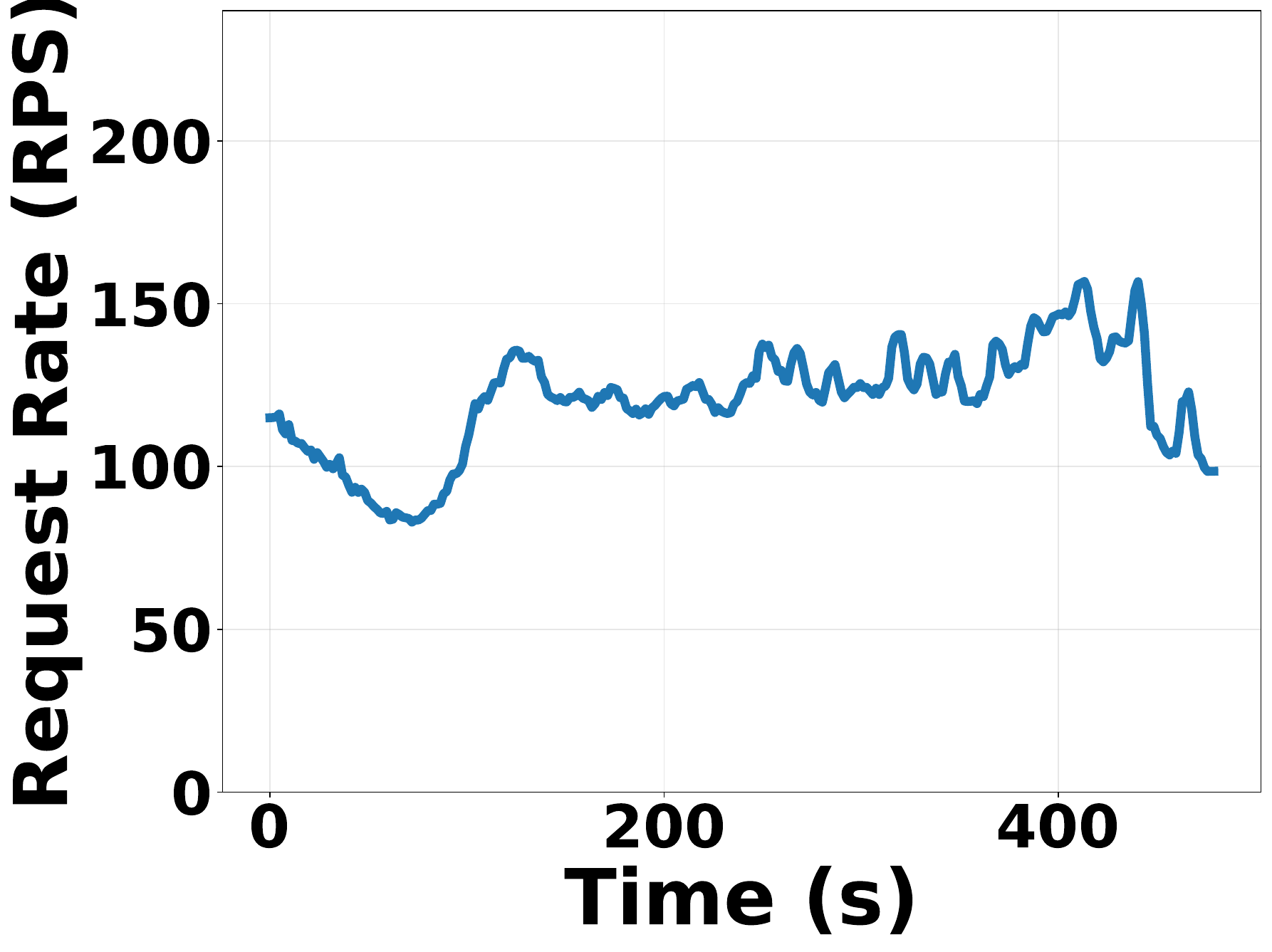}
    \caption{Smooth fluctuations}
    \label{fig:rps-1}
  \end{subfigure}
  \caption{Workload request rate (RPS) over time extracted from Alibaba Cluster Trace.}
  \label{fig:rps-combined}
\end{figure}

We generate workloads using data from the 2022 Alibaba Cluster Trace Dataset~\cite{alibaba2022}, a publicly available collection of real-world microservice traces from production environments. From this dataset, we select two representative microservices: one exhibiting smooth request rate fluctuations and another showing large, dynamic variations (as shown in Figure~\ref{fig:rps-combined}). These workloads are scaled to match our cluster capacity. To simulate real-time traffic, we use Locust~\cite{locust2025}, an open-source asynchronous load-testing tool, to drive traffic to the deployed services based on the selected traces.

\subsubsection{Baselines.} We compare our method against six baseline approaches: Linear Regression (LR), Multi-Layer Perceptron (MLP), Decision Tree (DT), Random Forest (RF), Sinan~\cite{zhang2021sinan}, and GRAF~\cite{park2024graph}. We exclude PERT-GNN~\cite{tam2023pert} from our comparison, as it focuses on API-level latency prediction without incorporating resource quota features, which are central to our window-level estimation approach.

\subsubsection{Evaluation Metrics.} We use three widely used metrics to evaluate the estimation accuracy of MSGAF and other models: Root Mean Square Error (RMSE), Mean Absolute Error (MAE), and Mean Absolute Percentage Error (MAPE). Lower RMSE, MAE, and MAPE scores indicate better estimation performance.

\subsection{Overall Performance} 


We evaluate model performance across varying latency conditions using the P50, P90, and P99 latency percentiles. The results are summarized in Table~\ref{tab:performance_comparison}, where our proposed MSGAF method consistently achieves the best performance across all metrics, benchmarks, and latency percentiles. The results reveal several key observations. Traditional linear and shallow models (e.g., Linear Regression, MLP) struggle to capture complex latency patterns, resulting in relatively high prediction errors. Tree-based models (e.g., Random Forest) perform better due to their ability to model non-linear feature interactions and hierarchical decision boundaries. The CNN-based method Sinan~\cite{zhang2021sinan} demonstrates strong performance by leveraging convolutional operations to extract local temporal patterns from system monitoring data, particularly excelling in higher percentile predictions. Meanwhile, the graph-based GRAF~\cite{park2024graph} incorporates service topology but fails to model multi-scale hierarchical relationships effectively, and lacks an adaptive mechanism to fuse features across different system granularities. In contrast, our MSGAF method integrates learnable multi-scale graph coarsening, adaptive fusion, and scene-aware estimation, enabling it to capture hierarchical system behaviors at the microscopic, mesoscopic, and macroscopic levels. The results demonstrate its consistently superior performance.

\begin{table*}[!t]
\centering
\caption{Evaluation of latency estimator methods across P50, P90, and P99 latency metrics. Bold indicates the best performance, underline represents the second best.}
\label{tab:performance_comparison}
\small 
\setlength{\tabcolsep}{5pt} 
\begin{tabular}{c|ccc|c|ccc}
\toprule
\multirow{2}{*}{\textbf{Method}} & \multicolumn{3}{c|}{\textbf{Online Boutique}} & \multirow{2}{*}{\textbf{Method}} & \multicolumn{3}{c}{\textbf{Sock Shop}} \\
\cmidrule(lr){2-4} \cmidrule(lr){6-8}
 & MAE & RMSE & MAPE &  & MAE & RMSE & MAPE \\
\midrule

\multicolumn{4}{c|}{\textbf{P50 (Median)}} & \multicolumn{4}{c}{\textbf{P50 (Median)}} \\
Linear & 50.83 & 65.96 & 38.90\% & Linear & 28.37 & 41.54 & 93.27\% \\
MLP & 21.83 & 31.84 & 13.78\% & MLP & 9.55 & 18.78 & 31.86\% \\
Decision Tree & 25.38 & 43.76 & 14.73\% & Decision Tree & 7.13 & 16.56 & 20.9\% \\
Random Forest & 22.11 & 34.93 & 13.06\% & Random Forest & \underline{6.38} & 13.26 & \underline{19.02\%} \\
Sinan & 20.04 & 30.97 & 12.32\% & Sinan & 6.49 & \textbf{12.93} & 30.74\% \\
GRAF & \underline{19.54} & \underline{30.79} & \underline{11.48\%} & GRAF & 11.19 & 21.46 & 32.39\% \\
 MSGAF &  \textbf{16.06} &  \textbf{25.76} &  \textbf{10.10\%} &  MSGAF &  \textbf{5.29} &  \underline{13.21} &  \textbf{13.57\%} \\
\midrule

\multicolumn{4}{c|}{\textbf{P90 (Tail)}} & \multicolumn{4}{c}{\textbf{P90 (Tail)}} \\
Linear & 95.78 & 130.65 & 35.61\% & Linear & 99.32 & 134.35 & 50.39\% \\
MLP & 55.88 & 87.39 & 17.86\% & MLP & 42.27 & 63.04 & 19.97\% \\
Decision Tree & 59.70 & 117.24 & 15.47\% & Decision Tree & 35.04 & 71.31 & 16.62\% \\
Random Forest & 54.03 & 104.96 & 14.16\% & Random Forest & 31.20 & 52.81 & 15.30\% \\
Sinan & \underline{45.72} & \underline{77.33} & 13.54\% & Sinan & \underline{22.16} & \textbf{36.93} & \underline{9.87\%} \\
GRAF & 52.18 & 90.77 & \underline{13.24\%} & GRAF & 43.35 & 64.50 & 22.15\% \\
 MSGAF &  \textbf{36.52} &  \textbf{67.02} &  \textbf{9.45\%} &  MSGAF &  \textbf{20.75} &  \underline{40.96} &  \textbf{9.06\%} \\
\midrule

\multicolumn{4}{c|}{\textbf{P99 (Extreme)}} & \multicolumn{4}{c}{\textbf{P99 (Extreme)}} \\
Linear & 201.44 & 329.32 & 32.66\% & Linear & 412.78 & 745.22 & 67.89\% \\
MLP & 163.25 & \underline{262.57} & 25.98\% & MLP & 184.41 & 329.68 & 29.06\% \\
Decision Tree & 176.66 & 355.67 & 26.18\% & Decision Tree & 194.08 & 417.40 & 35.57\% \\
Random Forest & 159.43 & 286.04 & 24.30\% & Random Forest & 159.33 & 299.66 & 29.96\% \\
Sinan & \underline{140.25} & 268.51 & \underline{21.54\%} & Sinan & \underline{119.18} & \textbf{205.21} & \underline{18.99\%} \\
GRAF & 155.53 & 289.60 & 21.84\% & GRAF & 197.22 & 333.91 & 32.67\% \\
 MSGAF & \textbf{132.92} & \textbf{260.86} &  \textbf{19.54\%} &  MSGAF &  \textbf{109.36} &  \underline{259.60} &  \textbf{17.63\%} \\
\bottomrule
\end{tabular}
\end{table*}

The analysis of different latency percentiles offers valuable insights into model behavior under varying system conditions. P50 latency, as a median metric, exhibits relatively stable and small fluctuations, resulting in narrow performance gaps across methods. For instance, on the Online Boutique dataset, the MAE difference between Random Forest and our approach is only 6.05, and as low as 1.09 on Sock Shop. This suggests that simpler models can perform adequately in capturing typical system behavior. In contrast, P99 latency, which reflects extreme system events such as anomalies or peak loads, proves highly volatile and unpredictable, posing significant challenges for all methods. Even with its enhanced modeling capabilities, our MSGAF approach experiences performance degradation under such extreme conditions, underscoring the inherent difficulty in predicting rare, high-latency events. P90 latency strikes a more favorable balance: it reflects the latency experienced by the majority of users without being overly influenced by extreme outliers, allowing all models to maintain reasonable accuracy. This makes P90 a more reliable and practical metric for latency-driven autoscaling decisions in real-world microservice environments.

\subsection{Ablation Study}

\begin{figure}[htbp]
    \centering
\includegraphics[width=0.96\columnwidth]{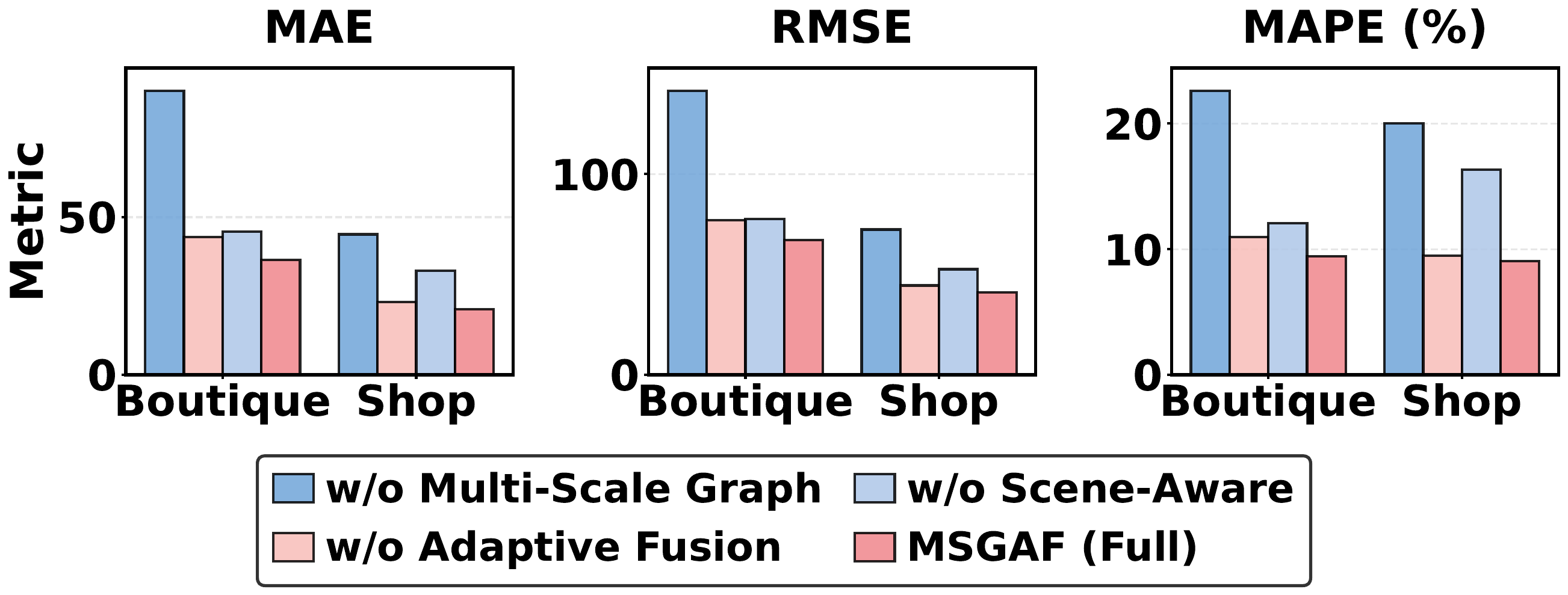}
    \caption{Ablation Study of MSGAF Components on Online Boutique and Sock Shop Benchmarks.}
    \label{fig:ablation-study}
\end{figure}

To investigate the contributions of different components of MSGAF, we conducted ablation studies under P90 percentile latency on Online Boutique and Sock Shop respectively. The results are presented in Figure~\ref{fig:ablation-study}. As expected, removing the multi-scale graph component results in the most significant performance degradation on both benchmarks, demonstrating that capturing hierarchical system behaviors across microscopic, mesoscopic, and macroscopic levels is crucial for accurate latency estimation. The cross-scale adaptive fusion shows moderate contribution with relatively smaller performance drops, while the scene-aware module exhibits varying importance across different system architectures, being particularly critical for Sock Shop. The ablation study confirms that all three components contribute meaningfully to the overall performance of MSGAF.

\subsection{Case Study: Hierarchical Level Analysis}

\begin{figure}[tbp]
  \centering
  \begin{subfigure}[b]{0.3\textwidth}
    \centering
    \includegraphics[width=\textwidth]{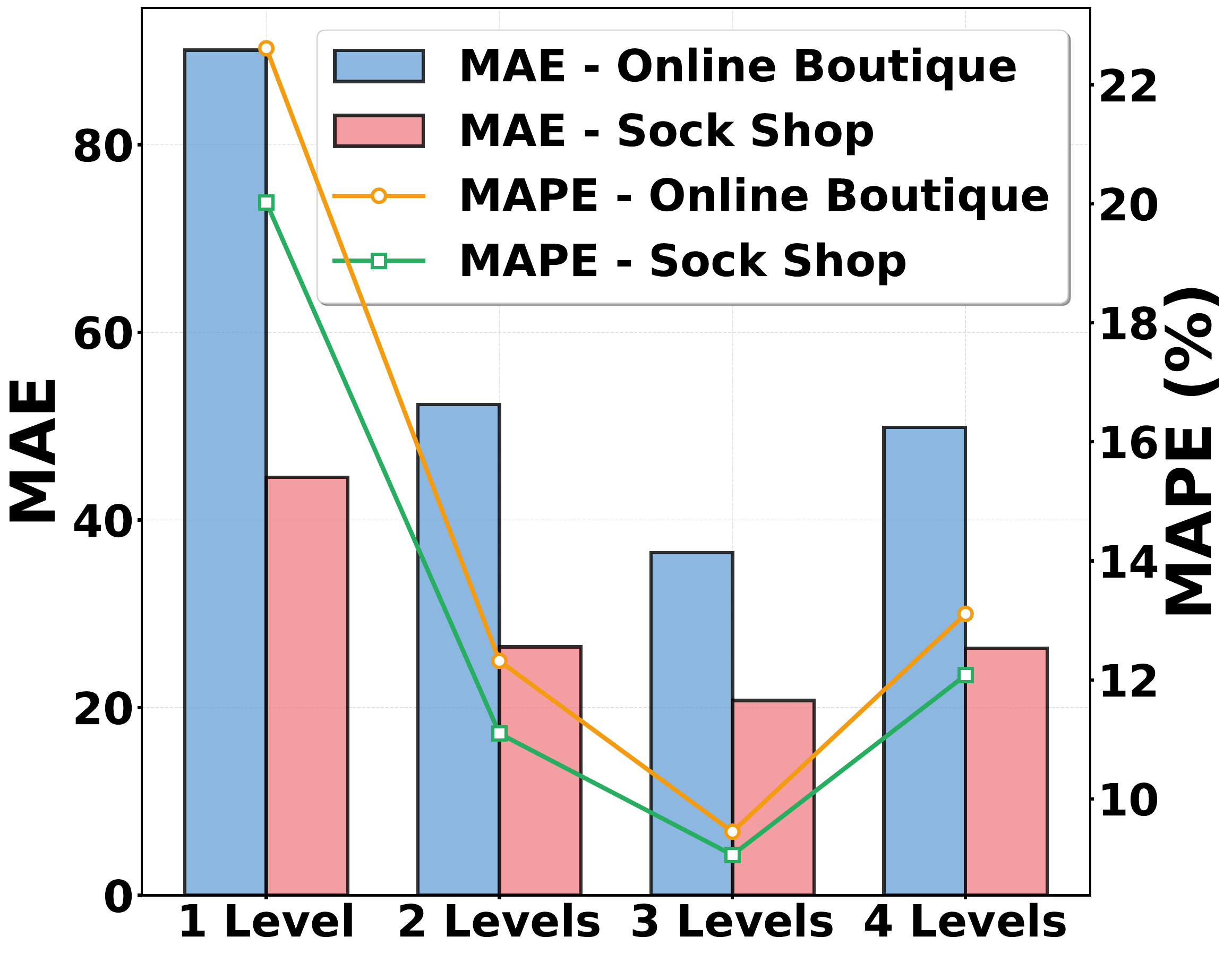}
    \caption{Performance comparison}
    \label{fig:mae_mape_analysis}
  \end{subfigure}
  \quad
  \begin{subfigure}[b]{0.3\textwidth}
    \centering
    \includegraphics[width=\textwidth]{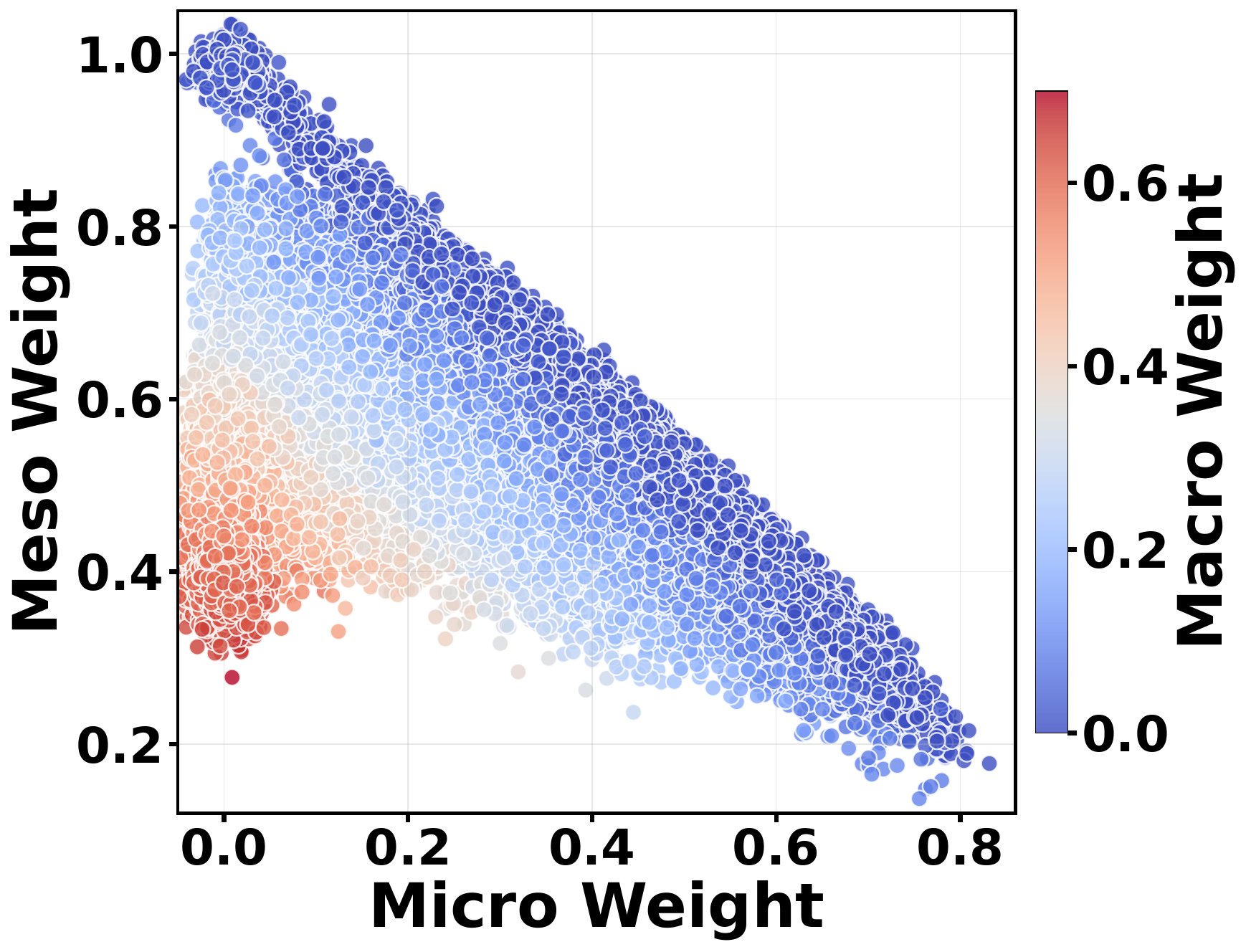}
    \caption{Fusion weights distribution}
    \label{fig:fusion}
  \end{subfigure}
  \caption{(a) Performance comparison in terms of MAE and MAPE across varying numbers of multi-scale graph hierarchy levels for the Online Boutique and Sock Shop benchmarks. (b) Distribution of fusion weights across multi-scale graph levels (Micro, Meso, Macro) under different operational scenarios on the Online Boutique benchmark under P90 latency estimation.}
\end{figure}

To further validate the effectiveness of our multi-scale graph architecture, we conducted an analysis of model performance under the P90 latency percentile, focusing on how the number of hierarchical levels affects prediction accuracy. The results, shown in Figure~\ref{fig:mae_mape_analysis}, compare model variants with 1, 2, 3, and 4 hierarchical levels across both the Online Boutique and Sock Shop benchmarks. The results demonstrate that the 3-level hierarchical configuration (micro, meso, and macro) achieves optimal performance on both benchmarks. Single-level models suffer from insufficient feature abstraction, while 2-level configurations lack adequate granularity to capture complex system behaviors. Adding a fourth hierarchical level leads to performance degradation, suggesting that excessive hierarchical granularity introduces noise and overfitting, diminishing the model's generalization capability. The 3-level configuration strikes an optimal balance between capturing multi-scale system behaviors and maintaining model tractability, validating our design choice for the hierarchical architecture. Beyond the optimal number of levels, the effectiveness of our approach also lies in its adaptive fusion capability. As shown in Figure~\ref{fig:fusion}, the weight distribution patterns across micro, meso, and macro scales dynamically adjust according to different workload scenes.

\section{Conclusion}
We propose MSGAF, a Multi-Scale Graph Adaptive Fusion framework with Scene-Aware Learning for microservice latency estimation in cloud-native environments. Our approach captures hierarchical system structures through learnable graph coarsening and dynamically adapts to varying workload conditions via expert networks with adaptive fusion. We also implement a non-intrusive monitoring system to support multi-granularity analysis. Experimental results show that MSGAF outperforms state-of-the-art methods, with ablation studies confirming the effectiveness of its core components.

\bibliographystyle{IEEEtran}
\bibliography{ref}


 





\end{document}